\documentclass[aps,preprint,showpacs]{revtex4}

\usepackage{epsfig}
\usepackage{latexsym}
\usepackage{amssymb}
\begin{document}
\title{Shear stress fluctuations in the granular liquid and solid phases}
\author{F. Dalton, F. Farrelly, A. Petri, L. Pietronero, L. Pitolli and G. Pontuale}

\affiliation{Consiglio Nazionale delle Ricerche, Istituto dei Sistemi Complessi, sede di Tor Vergata, Via del Fosso del Cavaliere 100, 00133 Roma, Italy.}

\date{\today}

\begin{abstract}

We report on experimentally observed shear stress fluctuations
in both granular solid and fluid states,
showing that they are non-Gaussian at low
shear rates, reflecting the predominance of correlated structures
(force chains) in the solid-like phase, which also exhibit finite
rigidity to shear. Peaks in the rigidity and the stress
distribution's skewness indicate that a change to the force-bearing
mechanism occurs at the transition to fluid behavior, which, it is shown, can be
predicted from the behavior of the stress at lower shear rates.

In the fluid state stress is
Gaussian distributed, suggesting that the central limit theorem
holds. The Fiber-Bundle Model with random load sharing effectively
reproduces the stress distribution at the yield point and also
exhibits the exponential stress distribution anticipated from extant
work on stress propagation in granular materials. \end{abstract}

\maketitle


Ubiquitous, Granular Media (GM) are important in many fields of
human activity and in the natural environment.  Their behavior can
often be described effectively by phenomenological laws, but
physical insight and intuition are often lacking
\cite{jaeger96,degennes99}. In particular, a generic friction law
for GMs would be a highly desirable achievement~\cite{lacombe00,robbins00}
both industrially
and in the research environment, with implications in other fields,
for example
earthquake dynamics~\cite{scholz-nature98}.  Acting against the
realization of such a law, however, are the fluctuations and
inhomogeneity innate to GMs~\cite{howell99}, which primarily arise from
the fact that GMs are non-ergodic, dissipative, out-of-equilibrium
systems: in dense GMs the kinetic energy can be many orders of
magnitude less than the work necessary to overcome gravitational
and friction forces, and so the dynamics is limited to the
vicinity of metastable, blocked states.  Moreover, stress
propagation is highly inhomogeneous and anisotropic
\cite{drescher72,howell99}.  When mechanically excited, therefore,
the mean behavior can be swamped by the fluctuations (in time,
space {\em and} ensemble) inherent to the system.  These
fluctuations carry important information on the microscopic
dynamics, and it seems therefore incumbent to investigate the
variance of physical quantities, by measuring not only their
averages but their full distributions. In fact, such information
may be of help in setting up theoretical frameworks or in
discriminating between theories with similar predictions for
macroscopic average quantities, for example granular friction.


In this Letter we report on the statistical properties of resistance
to shear in a dry granular medium.  It has long been
known~\cite{bagnold66} that, under increasing shear rate, a granular
medium displays a transition from a solid-like (stick-slip) to a
fluid-like (sliding) phase. However a complete understanding of this
transition seems yet unavailable~\cite{luding01}. Moreover, the
statistical properties of macroscopic quantities have never, to our
knowledge, been previously investigated in this context (previous
time-resolved measurements of stress were performed with different
aims in~\cite{miller96,nasuno97,nasuno98,hartley03,corcoran00} among
others). Our findings show that the two phases have different
statistical signatures unveiling a different origin for the internal
stress.


Our experimental set-up consists of a circular Couette cell
containing mono-disperse 2~mm glass beads.  An annular plate is
driven over the top surface of the channel by a motor {\em via} a
torsion spring. As the motor winds the spring, the torque on the
annular plate eventually exceeds the medium ``static friction''
$f_s$ and so the plate slips. It is similar to the system described
in Ref.~\cite{corcoran00}, the main difference being that the medium
in the apparatus presented here is free to select its own volume.
The inner and outer diameters of the channel are 20 and 36~cm
respectively and the channel is typically 5~cm deep.  To ensure a
granular shearing plane, the annular top plate has a layer of beads
glued to its lower surface, though, to avoid individual grains
jamming the system at the boundary, this layer of beads does not
extend to the full width of the channel. The system is initialized
before each experiment by pouring the medium into the channel, then
the system is run at a slow velocity ($\simeq 0.01$ rad/s) for a
long time (of order 1000 revolutions), such that it may approach a
stationary state (we have observed that the mean torque approaches a
steady state after the order of 10 -- 100 revolutions). During all
the tests reported here, the relative humidity was lower than
$40\%$.


Our measurements from the device consist of the angular position of
the annular top plate and the deflection of the torsion spring,
sampled at 1 kHz with an error~$\simeq 10^{-4}$ rad. Denoting
$\theta$ the angular coordinate of the plate in the reference frame
of the laboratory, the reaction torque $f_r$ ($ \le 0$) exerted by
the system during motion is
\begin{equation} \label{dynfriction} f_r = I \ddot{\theta} + \kappa (\theta -\omega_0 t),
\end{equation} where $I$ is the momentum of inertia of the plate, $\kappa$ is the spring
constant and $\omega_0$ the driving angular velocity. We identify
in particular the static torque, $f_s$, which is the maximum
torque achieved just before a slip, and a ``dynamic'' torque $f_d$
corresponding to trajectory points for which $\ddot{\theta}=0$.
The system displays proportionality between the average static
torque and the plate weight, as expected if the Mohr-Coulomb
failure criterion is to remain valid~\cite{savage84}.

\begin{figure}
\centering \epsfig{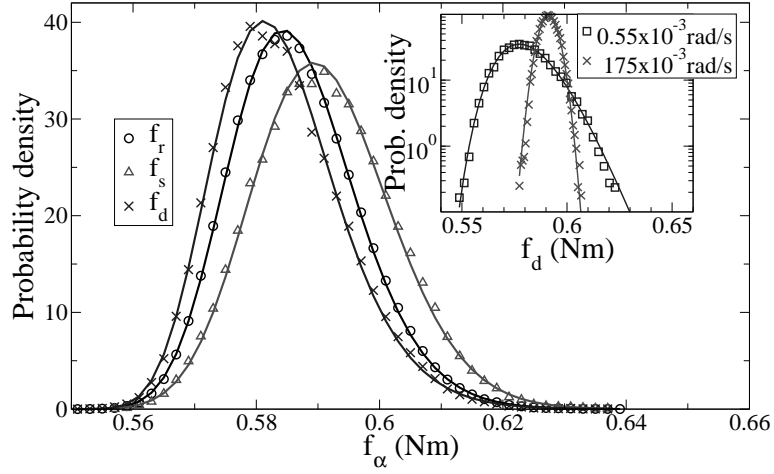}
\caption{\label{fig1} The
distribution of torque in the solid-like regime for $f_r$
({\large$\circ$}), $f_s$ ({\small$\triangle$}), $f_d$ ($\times$)
and their log-normal fits (lines, see Eq.~\ref{lognormal}). The
inset shows two sample experiments, one at slow driving
(solid-like regime) ($\Box$), one at fast  driving (liquid-like
regime) ($\times$), and their respective log-normal and Gaussian
fits, on semi-log axes.}
\end{figure}

We have investigated the statistics of the instantaneous
$f_{\alpha}$ ($\alpha=r,s,d$) for several different driving
velocities and employing springs of different constants.  It emerges
that at low driving velocities, i.e. in the solid-like, or
stick-slip regime, the distribution of $f_\alpha$ is not symmetric
and displays long tails at large values for all torsion springs
\cite{footnote}.
These distributions are shown in
figure~\ref{fig1} and, among all the curves tested, are best fitted
by a log-normal distribution (also shown, solid lines):
\begin{equation}
\label{lognormal}
p(f)=\frac{1}{\sqrt{2\pi}\sigma \cdot (f-f_0)} \exp \left[ -
\frac{(\ln(f-f_0)-\ln(\bar{f}-f_0))^2}{2\sigma^2} \right],
\end{equation}
with ($f > f_0 \forall f$).  It is worth
mentioning that the same shape has been found in experiments on
polymer films~\cite{briscoe85} and solid-on-solid
friction~\cite{johansen93}.  This suggests that, despite the
different nature of the three systems, similar mechanisms may lie
beneath the observed stick-slip behavior.

When the driving velocity is increased above a critical value
$\omega_c$, the fluctuations change: The inset to Fig.\ref{fig1}
shows the dynamic friction data $f_d$ in a comparison between a
slowly-driven ($\Box$) and a fast-driven ($+$) experiment and
their curve-fits (solid lines).  The fast-driven experiment is
clearly less asymmetric, and indeed the best-fit curve obtained
indicates a Gaussian fit.  From this feature we hypothesize that,
during the fluid phase, resistance to shear is dominated by short
lived, independent contacts and that the central limit theorem
applies (in the presence of dissipation this may not happen even
in a pure collisional regime~\cite{puglisi98}). Evidence for the
validity of this hypothesis will be given below.  On the contrary,
at low velocity, the stress bearing units are constituted by
correlated stress chains~\cite{drescher72,howell99}, for which
stress does not add independently.

\begin{figure}\centering
\epsfig{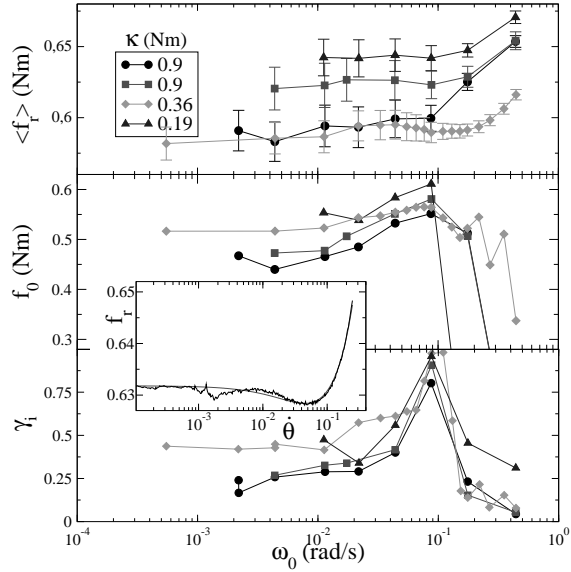}
\caption{\label{fig2}Fit parameters as functions of the driving velocity for different
springs. Top graph: mean torque with standard deviation bars; Center graph: rigidity
threshold; Bottom graph: distribution skewness.  At the transition, the skewness and the
rigidity peak then vanish, indicating that a symmetric Gaussian curve is a good model.
Inset: The instantaneous torque as a function of the instantaneous plate velocity at dynamic
friction points for a single experiment.}
\end{figure}

Investigation of the log-normal parameters as functions of the
driving velocity show that the average torque is approximately
constant at low velocity whereas it increases at high velocity, as
argued in previous work~\cite{savage84} and observed in other
experiments~\cite{ovarlez01}. Figure~\ref{fig2} shows the mean
torque $\langle f_r \rangle$, the rigidity $f_0$ (the minimum value
of the torque) and the
distribution skewness $\gamma$ as functions of the driving velocity
for different spring constants.  The small diversity in $<f_r>$
between the different torsion springs ($\sim 12\%$), with respect to
the change in the spring constant ($7.5$-fold increase), and the
absence of a clear trend indicate that the variation is due
primarily to configurational differences induced between different
series of experiments. This behavior is expected on a theoretical
basis and has been observed experimentally~\cite{bagnold66}. All
experiments show that at low velocity the width of the torque
distribution is also constant and decreases at high velocity while
at the same value of $\omega_0$, $\omega_c \simeq 10^{-1}$ rad/sec,
the rigidity and the distribution skewness peak before vanishing.
Thus the distribution shape crosses to a symmetric Gaussian, which
is the limit form of the log-normal when $\sigma\ll 1$:
\[
\frac{1}{\sigma^2}\left[\ln(f-f_0)-\ln(\bar f-f_0)\right]^2\simeq \left[\frac{\bar f-f}
{(\bar f-f_0)\sigma}\right]^2. \]
This is the fluidization transition, above which the plate continuously slides.

We have also measured the dependence of the instantaneous torque on the instantaneous plate velocity
$\dot{\theta}$. This is shown at the dynamic friction points in the inset to Fig.~\ref{fig2} for a
representative experiment.
The data show a roughly constant value of the torque for $2$ decades
of the plate velocity, as in solid friction. As the latter
approaches the fluidization value $\omega_c$, there is a slow
drop followed by a somewhat faster increase, as predicted theoretically~\cite{jaeger91}.
Such behavior has also been observed in some experiments~\cite{lubert01}
on GMs and model systems~\cite{heslot94} and, in Fig.~\ref{fig2} has been fitted by
a curve of the form:
\begin{equation}\label{eq-friction}
f_d(\dot{\theta})  =
f_0+\gamma(\dot\theta-2\omega_f\ln(1+\dot\theta/\omega_f)).
\end{equation}
with $f_0=0.63$ Nm, $\gamma=0.19$ Nms/rad, $\omega_f=0.043$ rad/s.
Finally, when the velocity becomes comparable or larger than the
critical driving velocity $\omega_c \simeq 10^{-1}$ rad/sec, the
torque increases almost quadratically ($f_d \approx
\dot{\theta}^{1.9}$ in the figure).
From this fact two conclusions can be drawn: \\

{\em i)}: the medium is almost fully fluidized above $\omega_c$.
Only in this
case is resistance to stress expected to increase
quadratically with the shear rate \cite{savage84}, In this regime
dynamics is dominated by collisions, and from the Gaussian
distribution observed we argue that conditions for applying the
central limit theorem hold. Moreover, the plate velocity in our
experiments can be identified with the shear strain rate up to a
proportionality factor.

{\em ii)}: fluidization occurs intermittently whenever the plate's
instantaneous velocity is larger than the critical value, even at
low driving. Thus the value of the fluidization velocity can be
predicted while driving the system at a much lower velocity.  Note,
however, that not all slip events reach the high velocity necessary
for the fluid state, and so we deduce that for these events, the
solid-like phase remains dominant. As for the mean stress as a
function of the driving velocity, here there is no
evidence of a trend with respect to the spring constant.\\


The stress distribution observed in the structured (solid-like)
phase, Eq.~(\ref{lognormal}), can be reproduced in a robust way by
modeling the evolution of stress within the medium as occurring
along a finite number of interacting force chains in a stress field.
 Before describing the model in detail however, it is
necessary to understand precisely the behavior which the model must
exhibit.  In GMs, both compressional and shearing stress propagates
in the medium through highly directed chains of
grains~\cite{drescher72,howell99} that are topologically
one-dimensional, although branching and crossing become highly
probable when the packing fraction
increases~\cite{howell99,makse00}.  These chains form a random
network which appears rather rarefied, screening most of the medium
volume from stress.  The stress sharing among the supporting
structures has been measured in granular media in
two~\cite{howell99,behringer99} and three dimensions~\cite{makse00},
and has been found to display a long exponential tail for stresses
$s$ above the mean value, $\bar s$. That is, the number of stress
chains supporting a stress $s$ is proportional to
\begin{equation}
\label{stressdistr} n(s)|_{s>\bar s}\sim \exp[-s/\bar s].
\end{equation}
This behavior seems quite general, since it has been also observed
in a compressed emulsion~\cite{brujic03} and in numerical
simulations of glasses~\cite{hern01}. Thus the model should {\em
also} satisfy the constraint of reproducing in a robust way
an exponential tail,
Eq.~(\ref{stressdistr}) in the distribution of the chain loads
before the final yielding, besides  the log-normal shape of the
yield-stress
distribution.

To adequately model the stress evolution of the force chains within
the medium, perhaps the simplest non-trivial scheme is to consider a
bundle of one-dimensional uniform fibers supporting a longitudinal
stress. The first such Fiber-Bundle Model (FBM) was
introduced~\cite{peirce26} and extensively
investigated~\cite{daniels45,suh70} for describing the fracture of
textiles. It consists of a bundle of $N$ fibers loaded in parallel,
whose individual strength to failure is randomly extracted from some
preselected distribution $p(\sigma)$.  The present analogy
associates a stress chain in the medium with a fiber, and has also been
employed for describing the constitutive behavior of compressed
GMs~\cite{hidalgo02}. Besides the load sharing law among fibers, we
are interested in the distribution of the global failure stress $S$
when the number of fibers $N$ is not large, as results from imaging
of force chains would
suggest~\cite{howell99,drescher72,behringer99}.  This corresponds to
the distribution of the yield stress of the medium under the
increasing elastic shear.

In the primitive FBM the re-distribution of the loads is global,
also known as "democratic sharing". That is, as soon as the
increasing applied stress exceeds the failure load of one (the
weakest) fiber, this breaks and its load is evenly redistributed
among all the remaining fibers. If it happens to some of them to be
in turn overloaded, they too break and redistribute the load to all
the others, and the cycle is repeated until no more fibers break.
Then the stress on the system is further increased until the
following fiber breaks, and so on. The FBM adopted here assumes
instead that whenever a fiber fails its load is re-distributed in
random proportions $p\in [0,1]$ and $q=1-p$ among two other fibers
chosen randomly from the remaining fibers (Random FBM). In fact the
formation of stress supporting chains and the exponential tail in
the stress sharing (Eq.~(\ref{stressdistr})) are explained for
compressional stress by  a simple mechanism (commonly known as the
$q$-model~\cite{liu95}) in which the stress exerted on a grain is
randomly transmitted to some neighboring
grains~\cite{liu95,socolar98,marconi00,claudin98}.

\begin{figure}\centering
\epsfig{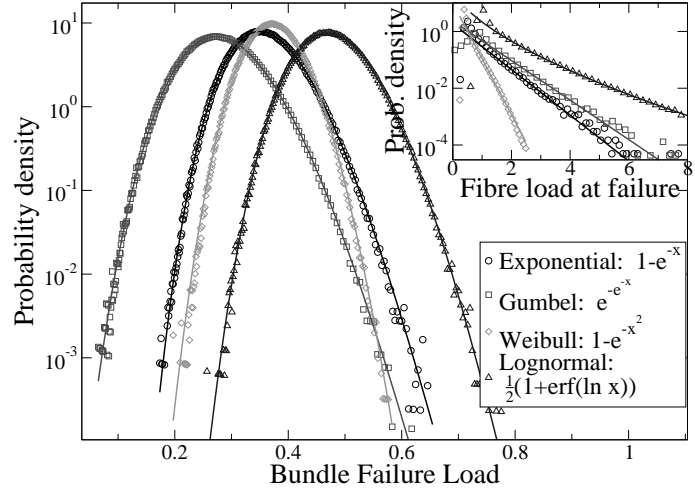}
\caption{\label{fig3}
The bundle strength for a selection of initial fiber distributions with a log-normal fit to
each (lines).  The inset demonstrates the exponential tail (stretched exponential in the
log-normal case) in the distribution of fiber-loads at failure for these same initial
distributions.  The key to the graph indicates the initial cumulative probability
distributions used to create the results (the number of fibers $N=100$).}
\end{figure}

Generically speaking, FBMs with $N \to \infty$ possess rupture at a
pre-determined value $S_p$ set by the initial distribution of fiber
thresholds, whereas with $N$ not large, $S_p$ assumes an asymmetric
distribution for a large class of initial fiber strength
distributions $p(\sigma)$ \cite{daniels45,suh70}. We have performed
several simulations exploring the parameter space of the RFBM;
principally how different forms for the probability distribution of
the fiber strengths affects the behavior of the model.  In
Fig.~\ref{fig3} the distribution of bundle strengths for a range of
initial fiber strength distributions is shown. The result is
observed to closely follow the log-normal distribution for a wide
variety of input distributions (results for exponential, Gumbel,
Weibull and log-normal inputs are shown, but similar results are
obtained by other distributions $p(\sigma)$).  The corresponding
fiber load distributions just before the final rupture are shown in
the inset to Fig.~\ref{fig3}. The final break-down corresponds to
the system yield and triggers a slip event.

In all cases the load sharing is characterized by a long exponential
tail which, simulations show, is quickly set up after starting to
load the system~\cite{fbm-work}. For exponential, Gumbel and Weibull
fiber strengths (amongst others), the fiber-load distribution in
RFBM seems to have a robust exponential tail, while for log-normal a
stretched exponential is obtained, though with appropriate
parameters the stretching exponent $\beta\simeq 1$. If we assume
that the overall strength of a force chain would be determined by
its weakest element, then the limit distributions for extreme
values~\cite{galambos} would be natural choices for the initial
fiber strengths distributions (a detailed study of the model with
quantitative reconstruction techniques of the stress distribution
~\cite{tagliani03} will be published elsewhere~\cite{fbm-work}).

Thus, when loaded towards its stability limits, RFBM yields both
the internal load sharing, Eq.~(\ref{stressdistr}), and the yield
stress distribution, Eq.~(\ref{lognormal}), observed in the
experiments. It can also in principle consider the formation and
destruction of force fibres during shear motion, by including the
possibility that fibers ''heal''~\cite{selinger91}, in such a way
that in the stick slip regime the system is constantly at the edge
of failure, constituting a possible starting point for
understanding the similarities observed in the stress
distributions in solid-on-solid and polymer film shear
experiments. Our results indicate that in the solid regime the
system self-organizes in structures, or chains, supporting the
stress with an asymmetric distribution about the mean value which
can be robustly reproduced by the Fiber-Bundle model with random
load sharing. By increasing the driving velocity instantaneous
fluidization becomes more and more frequent until the fluid regime
is reached, in which the dynamics is collisional and stress
distribution is symmetric. We hope that the knowledge of stress
fluctuation will be of help in improving the theoretical
description of the granular solid-liquid transition.

We are grateful to A. Baldassarri and S. Zapperi for pointing out the logarithmic drop of instantaneous friction with velocity.
This work was supported by the EU Contract No. ERBFMRXCT980183 and by the FIRB Project RBAU01883Z.

\bibliographystyle{apsrev}
\bibliography{sheargrbibv2,fbmbibv2}

\begin{thebibliography}{40}
\bibitem[{foo()}]{footnote}
\bibinfo{note}{Very recently \cite{baran04} a system with exactly the same
  geometry of the experiment presented here has been simulated for the first
  time. Even in that case the tangential stress distribution has been found to
  be asymmetric in the solid phase. However a quantitative comparison with the
  experimental results is difficult since the system is simulated in the
  absence of gravity and statistical distributions refer to both time and
  spatial fluctuations.}
\end{thebibliography}

\newpage

\end{document}